\documentclass[12pt]{article}
\usepackage{color}
\usepackage{amssymb,amsmath}

\usepackage{color}

\numberwithin{equation}{section}

\begin{document}
\title{The action for higher spin black holes in three dimensions}
\author{M\'{a}ximo Ba\~{n}ados$^{a}$, Rodrigo Canto$^{a}$ and Stefan Theisen$^{b}$  \\
{\small $^a$Departamento de F\'{\i}sica, P. Universidad Cat\'{o}lica de Chile, Santiago 22, Chile}\\
{\small $^{b}$ Institut f\"ur Theoretische Physik, Universit\"at Heidelberg, 69120 Heidelberg, Germany\footnote{permanent address:
Max-Planck-Institut f\"ur Gravitationsphysik (Albert-Einstein-Institut), 14476 Golm, Germany}}}

\maketitle

\begin{abstract}

In the context of (2+1)--dimensional Chern-Simons $SL(N,\mathbb{R})\times SL(N,\mathbb{R})$ gauge fields and spin N black holes we compute the on-shell action and show that it generates sensible and consistent thermodynamics. In particular, the Chern-Simons action solves the integrability conditions recently considered in the literature.

\end{abstract}

\setcounter{equation}{0}
\section{Introduction}

Three dimensional gravity with a negative cosmological constant is an interesting arena for a variety of question. It is much simpler than higher dimensional gravity while still being non-trivial. Topologically massive gravity\cite{Deser:1981wh}, the construction of two copies of the Virasoro algebra as the asymptotic symmetry algebra\cite{BH}, and the existence of black hole solutions \cite{BTZ,BHTZ}, are just a few examples. Noteworthy, the Brown-Henneaux symmetry may be viewed as a precursor of the AdS/CFT correspondence.

Gravity in three dimensions does not contain propagating degrees of freedom. It is a topological theory and it can also be formulated as a $SL(2,\mathbb{R})\times SL(2,\mathbb{R})$ Chern-Simons (CS) theory. This alternative formulation does not use a metric, but rather a gauge field. The integer level $k$ of the CS-theory, the AdS-radius $\ell$ and Newton's constant $G$ and the central charge of the
two copies of the Virasoro algebra are related via

\begin{equation}
k={\ell\over 4 G}={c\over 6} \,.
\end{equation}

While in the metric formulation the solutions to the
equations of motion are Einstein mani\-folds with negative cosmological constant, they
are flat connections in the CS formulation. The Chern-Simons formulation allows for an immediate generalization to higher rank gauge groups. This raises two obvious questions. (i) what is the field-content in terms of metric-like fields and (ii) what is the asymptotic symmetry algebra? These questions were addressed and answered in \cite{Henneaux:2010xg,Campoleoni:2010zq} where it was found that
e.g. for $SL(3,\mathbb{R})\times SL(3,\mathbb{R})$ with principally embedded
$SL(2)\hookrightarrow SL(3)$, the theory contains gravity and a spin-3 field -- both with no propagating degrees of freedom. The symmetry algebra was shown
to be two copies of the Zamolodchikov $W_3$ algebra \cite{Zamolodchikov:1985wn}
with the same central charge as
in the pure gravity theory. While a formulation of the dynamics of these
higher spin theories in terms of metric-like fields is still unknown, ref. \cite{Campoleoni:2010zq} proposed a way to construct the metric and the spin-three field from the $SL(3)$ connection.

Several generalizations have been discussed in the literature, e.g. higher rank gauge
groups and other embeddings \cite{Campoleoni:2011hg} and super-symmetrization \cite{Henneaux:2012ny}, to name just two. Another interesting question which one might ask, is whether there is a generalization of the BTZ black hole solution which carries $W$-charge. Such black hole
solutions were indeed constructed and their properties were discussed in \cite{Gutperle:2011kf,Ammon:2011nk,Kraus:2011ds,Castro:2011fm}.

One of the conditions which were imposed on the
prospective BH solutions was that they should possess a smooth BTZ limit. Furthermore,
it was required that they are described by two pairs of conjugate thermodynamical
variables which can be used to formulate a sensible BH thermodynamics.
Their thermodynamics has been analyzed by using integrability conditions that follow
from the assumption of the existence of a partition function. It is the purpose of this
note to show that the partition function indeed exists and that it can be computed in a
straightforward way from the Chern-Simons action.

In section 2 we discuss the Chern-Simons action and its variation in general terms, with
particular emphasis on boundary terms. In Section 3 we construct the action for spin $N$ fields
and exhibit its canonical structure.
In section 4 we treat the case $N=3$ as a particular example and in section 5 we show the relation
with the metric formulation for this case. A generalization for arbitrary $N$ has not been
worked out yet. Finally, in an appendix we explain the relation between the parametrization used
here with the one which is used when one discusses the asymptotic symmetries and charges.

\section{The Chern-Simons action and its variations}
\label{CSACTION}

The relation between $SL(3)\times SL(3)$ Chern-Simons theory and spin three fields has already been discussed.  The problem we face is to build a black hole with consistent thermodynamical properties.
This question can and will be discussed for the general case of $SL(N)\times SL(N)$ CS theory
with principal embedding $SL(2)\times SL(2)\hookrightarrow SL(N)\times SL(N)$ which
leads to a theory with spin $j=2,\dots, N$ fields. Other embeddings can presumably be
discussed along the same lines, but we will not attempt to do so.

We work on the solid torus, which is the topology of the Euclidean black hole, having one contractible and one non-contractible cycle. The contractible cycle is parameterized by the coordinate  $t$ and the non-contractible one by $\phi$. The ranges of the coordinates will be fixed throughout the paper,
\begin{equation}\label{ranges}
0\leq t <1, \ \ \ \ \ 0\leq \phi < 2\pi.
\end{equation}

It is customary to consider the range $1\leq t < \beta$. Equivalently, one can set $0\leq t<1$ by a coordinate reparametrization. The price to pay is that $\beta$ now appears in the field. (For the Schwarzschild metric,  the $tt$ metric component of the metric reads $g_{tt}=-\beta^2(1-2m/r)$.) When dealing with the action, it is convenient to have all varied parameters in the fields. In this way, the integration limits in the action are fixed.

We also introduce the
chiral coordinates $z,\bar z$
\begin{equation}\label{coords}
z= t + \phi, \ \ \ \ \bar z = t - \phi.
\end{equation}
In the Euclidean sector $t$ is replaced by $it$. The relation between the connection on both
set of coordinates is,\footnote{We will denote the
connections of the second $SL(N)$ factor by $\bar A$, but since the problem is
symmetric in the two factors, we will mostly restrict to $A$.}
\begin{equation}\label{At}
A_t = A_{z} + A_{\bar z}, \ \ \ \ \   A_\phi = A_{z} - A_{\bar z}.
\end{equation}

Finally, a radial coordinate $r$ runs from infinity (a torus) to the horizon defined as the ring at the center of the solid torus. However, as shall see, the radial coordinate will play no role in our analysis.

Our main goal is to evaluate the Chern-Simons action (wedge products are omitted),
\begin{equation}\label{ICS}
I_{0}[A] = {k \over 4\pi} \int \mbox{Tr} \left(  AdA + {2 \over 3} A^3 \right)
\end{equation}
for a given configuration $A=A_{\mu} dx^\mu$ satisfying the equations of motion
\begin{equation}\label{}
F=0.
\end{equation}
We have included the subscript $0$ to stress the fact that this action may need the addition of boundary terms to fulfill the desired boundary conditions.

Recently\cite{Gutperle:2011kf}, in the context of higher spins fields, a shortcut has been used to evaluate the action. As an example, consider the Schwarzschild black hole characterized by total energy $M$ and inverse temperature $\beta$. Regularity implies $\beta=8\pi M$.
Writing
\begin{equation}\label{Mbeta}
M = {\partial I[\beta] \over\partial \beta}
\end{equation}
and replacing in the regularity condition one finds a differential equation for $I[\beta]$ whose solution is $I = {\beta^2 \over 16\pi}$. This coincides exactly with the (regulated) Einstein-Hilbert action evaluated on the Schwarzschild  solution\cite{Gibbons:1976ue}.  If there are more parameters (rotating or charged black holes), the regularity conditions imply certain integrability conditions allowing the existence of an action. Again, its value can be found by integrating the regularity conditions\cite{Gutperle:2011kf}.

At this point one may wonder why should one care about evaluating the action directly.
The above method seems to provide the right value in a rather quick way.
There are two reasons. First, the notion of conjugate variables is contained in the action. For the Schwarzschild black hole, the action dictates that $\beta$ and $M$ are conjugate, and justifies (\ref{Mbeta}). Second, the above trick is valid only semiclassically. If one is interested in quantum corrections, the actual action is unavoidable.

We now turn to the evaluation of the Chern-Simons action. Actually, before evaluating the action itself, we consider its variation which contains the information of canonical coordinates. Varying (\ref{ICS}) we obtain,
\begin{eqnarray}
  \delta I_0[A] &=& {k \over 4\pi} \int_{\mathbb{R}\times T_2} \mbox{Tr}( F\wedge \delta A) + {k \over 4\pi} \int_{T_2}  \mbox{Tr} ( A \wedge \delta A) \nonumber\\
    &=& 0 \ \  + \ \  {k \over 4\pi} \int_{r\rightarrow \infty} dt d\phi\, \mbox{Tr}  (A_\phi \delta A_t - A_t \delta A_\phi)\, .\label{dI0c}
\end{eqnarray}
In this calculation, the bulk part is zero as a consequence of the equations of motion. The boundary term is evaluated for large $r$ although, as we shall see, the radial coordinate is largely irrelevant.
The variation (\ref{dI0c}) is important to understand how to compute the free energy. Different boundary conditions require different boundary terms, and this information is encoded in (\ref{dI0c}), not in the actual value of the action on a given solution. A possible set of boundary conditions are the chiral fields $A_t = \pm A_\phi$. The boundary term in (\ref{dI0c}) is zero and the action has well-defined variations. The resulting asymptotic algebra is the affine KM symmetry (one dimensional gauge transformations). However, this set of boundary conditions are not enough to ensure AdS asymptotics. As first discussed in \cite{Coussaert:1995zp} in the context of the $SL(2,\mathbb{R})\times SL(2,\mathbb{R})$ theory, extra conditions on the currents are necessary to find the conformal symmetry with the correct central charge. For the higher spin theory, the extra conditions leading to the W symmetry were discussed in \cite{Henneaux:2010xg,Campoleoni:2010zq}.

Note that  a boundary term at the horizon (center of the solid torus) does not appear   because (\ref{ICS}) is a covariant action.  There is a key difference between the covariant and Hamiltonian actions. The Hamiltonian action is built using a time foliation which, for black holes, breaks down at the horizon. To make sense of the action and its variation one removes a small disk around the horizon (introducing an artificial boundary), imposes appropriate boundary conditions and one then lets the radius of the disk go to zero. The result of this procedure is an extra boundary term which turns out to be equal to the entropy. See \cite{Hawking:1980gf,Teitelboim:1994az} for discussions on this point. On the other hand,  the covariant action has no coordinate issues. The only boundary is the circle$\times$time at infinity.

We now turn to the problem of evaluating the action itself. Since on-shell $F=0$, one may write $dA = -AA$ and conclude that the action for a field satisfying the equations of motion is proportional to $-{1 \over 3} \int A^3$. This is a volume integral extending all the way to the horizon. The formula $-{1 \over 3} \int A^3$ gives the right value provided one writes the black hole field in regular coordinates (e.g. Kruskal coordinates) at the horizon. Otherwise, the result may be incorrect. For example, if one plugs the black hole field written in Schwarzschild coordinates one obtains $\int A^3=0$, which is not the right result.

Fortunately, in 2+1 dimensions, Kruskal coordinates can be avoided because there exists an alternative way to evaluate the action which is  `half way' between the covariant and Hamiltonian actions. The idea is to use angular quantization\cite{Banados:1998ys}. Let us foliate the solid torus by disks of constant $\phi$. This foliation is regular everywhere without any subtleties at the horizon. Making the angular 2+1 decomposition
\begin{equation}\label{}
A  = A_\alpha dx^\alpha + A_\phi d\phi\,,
\end{equation}
where $x^{\alpha}$ are coordinates on the disk, the action (\ref{ICS}) becomes, keeping track of all boundary contributions,
\begin{equation}
  I_{0,\rm angular} =  {k \over 4\pi}\int d\phi \int d^2x \mbox{Tr} \epsilon^{\alpha\beta} (- A_\alpha \partial_\phi A_\beta + A_\phi F_{\alpha\beta} ) - {k \over 4\pi}\int_{r\rightarrow \infty} dtd\phi\, \mbox{Tr} (A_t A_\phi)\,. \label{af}
\end{equation}
The bulk piece of this action is explicitly covariant with respect to the 2-dimensional leaves of the foliation. One can use Cartesian or any other regular set of coordinates to parameterize the disk. At the outer boundary, on the other hand, the Schwarzschild $t,r,\phi$ coordinates are regular and we can evaluate the boundary term using them. For spherically symmetric on-shell fields, the bulk piece is zero and we only need to take care of the boundary term at infinity. The on-shell value of the Chern-Simons action is therefore
\begin{equation}\label{Ios}
I^{\rm on-shell}_{0} = -{k \over 4\pi}\int_{r\rightarrow \infty} dtd\phi\, \mbox{Tr} (A_t A_\phi)\,.
\end{equation}
We can evaluate this integral on the black hole field by simply plugging it into \eqref{Ios}, e.g., in Schwarzschild coordinates. Besides its simplicity, (\ref{Ios}) has the virtue of being finite without requiring any regularization (see below).

The expression (\ref{Ios}) is the on-shell value of (\ref{ICS}).  However, it is not the end of the problem. Before plugging a solution into (\ref{Ios}), we need to make sure that (\ref{ICS}) actually has an extremum on that solution. Otherwise
we will get the wrong value for the action.
The information about the extrema is encoded in (\ref{dI0c}), not in (\ref{Ios}).

The general structure of the problem is familiar.
For a given set of solutions parameterized by $2n$ constants $z^a=\{m,q,\beta,\mu...\}$ (mass, charge, inverse temperature, chemical potentials,  etc.), the on-shell variation of the action is a pure boundary term
\begin{equation}\label{dz}
\delta I_0 = \sum_{a=1}^{2n}\ell_a(z) \delta z^a.
\end{equation}
Consistency of the variational principle does not allow to fix all $2n$ parameters at the outer boundary, but only half one them,
which we denote by $q^i$. Once their boundary data are chosen, Darboux' theorem states that (\ref{dz}) can be reorganized in the form
\begin{equation}\label{C}
\delta I_0 = \sum_{i=1}^n p_i \delta q^i + \delta C(p_i,q^j)\,,
\end{equation}
where $p_i$ are the conjugate momenta and $C(p,q)$ is, in general, a non-trivial function.

From (\ref{C}) we see that the correct action for this choice of boundary data is
\begin{equation}\label{i1}
I_1 = I_0 - C,
\end{equation}
which satisfies
\begin{equation}\label{C1}
\delta I_1 = \sum_{i=1}^n p_i \delta q^i\,,
\end{equation}
$I_1$ is extremal when the $q$'s are held fixed.  Other choices for the set of free parameters
lead to other actions which are related to $I_1$ by a Legendre transformation.

The solutions characterized by the $2n$ parameters $p_i$ and $q^i$ make perfect sense asymptotically, but are not all regular in the interior because, as in the case of the black hole, there might not be a boundary.
This means that one cannot arbitrarily chose $2n$ constants, but only half of them. The other half is determined by $n$ `regularity conditions',
\begin{equation}\label{chi}
\chi_l(p_i,q^j)=0, \ \ \ \ \ l=1,\dots,n\,.
\end{equation}
For example, for the Schwarzschild black hole, the parameters are $\beta$ and $m$. Asymptotically they are independent. However, only solutions with  $\beta - 8\pi m=0$ can be extended all the way to the horizon. The regularity conditions allow us to express all $p$'s as functions of the $q$'s,
\begin{equation}\label{}
p_i = p_i(q^j).
\end{equation}
With these formulae at hand we can now express $I_1$ as a function of the $q$'s, $I_1 = I_1[q,p(q)]$.   If the regularity conditions and action have been identified correctly then the functional $I_1[q]$ must satisfy,
\begin{equation}\label{dqI}
{\partial I_1[q,p(q)] \over\partial q^j} = p_j\,,
\end{equation}
consistently with (\ref{C1}). This provides a final check of the procedure. Below we shall apply this
to the on-shell Chern-Simons action for higher spin black holes, proving explicitly the consistency of (\ref{dI0c}) and (\ref{Ios}).

Before doing this in the next section,
we briefly consider the more familar time foliation. This leads to the
Hamiltonian action and the issue of a boundary term at the horizon, which one expects to be related to the entropy. In a time foliation $A=A_idx^i + A_t dt$ the Chern-Simons action becomes, keeping track of boundary contributions,
\begin{equation}\label{tf}
  I_{\rm Hamiltonian} =  {k \over 4\pi}\int dt \int dr d\phi ( A_r \partial_t A_\phi + A_t F_{r\phi} )
+{k\over 4\pi}\left\lbrace\int_{\infty} dt d\phi\, \mbox{Tr}(A_t A_\phi) - B_+\right\rbrace\,.\\
\end{equation}
This is a Hamiltonian action in the sense that the equations are
first order in time derivatives, $A_r$ and $A_\phi$ are canonically conjugated, and $A_t$ is the Lagrange multiplier enforcing the constraint $F_{r\phi}=0$.

As discussed before, since the leaves of the time foliation all meet at $r=0$ where the foliation is singular, one needs to add a boundary term at the horizon, $B_+$ in (\ref{tf}). This boundary term is absent in the angular foliation, as it is regular at $r=0$. For metric theories of gravity, this term has been extensively discussed in the literature. To our knowledge, similar results in the connection representation are not known.

The boundary terms appearing in (\ref{tf}) are {\it not}, in principle, appropriate for a canonical, microcanonical, or any other choice of variation. These terms are simply contributions that arise when performing the 2+1 decomposition. In other words, the action (\ref{tf}) is still exactly the same as the action (\ref{ICS}) and also as the action \eqref{af}.
Expressions (\ref{af}) and (\ref{tf}) are simply different ways to rewrite (\ref{ICS}).
We shall need extra boundary contributions (see below) to build the canonical or microcanonical actions.

Comparing the on-shell values of \eqref{af} and \eqref{tf}, we can evaluate the boundary term at the horizon,
$B_+$.
From the point of view of the torus at infinity, the time and angular foliations differ only by an orientation sign. This explains the opposite signs of the boundary term at infinity in (\ref{tf}) and (\ref{af}).
The bulk pieces in (\ref{af}) and (\ref{tf}) are zero for stationary, spherically symmetric on-shell fields.
The only contributions to the actions come from their boundary terms.
We then conclude
\begin{equation}\label{smarr}
-\int_{\infty} dt d\phi\, \mbox{Tr}(A_t A_\phi) =  \int_{\infty} dt d\phi\, \mbox{Tr}(A_t A_\phi) - B_+
\end{equation}
or,
\begin{equation}\label{B+}
B_+ = 2 \int_{\infty} dt d\phi\, \mbox{Tr}(A_t A_\phi).
\end{equation}
We have thus found a formula for the boundary term at the horizon in the Hamiltonian action, expressed as an integral at infinity. We mention that for the BTZ black hole, relation  (\ref{smarr}) becomes equivalent to the Smarr relation $\beta(M+\Omega J) = -\beta(M+\Omega J) +  S$. See \cite{Banados:1998ys} for more details on this case.

\section{The action for spin $N$ fields}
\label{SPINN}

Our main goal in this paper is to point out that the (on-shell) Chern-Simons action provides sensible thermodynamics and, in particular, solves the integrability conditions discussed in \cite{Gutperle:2011kf}. All this is expected because flat connections are solutions to the Chern-Simons equations of motion. To our knowledge, however, this connection has not been made explicit in the literature.
We now turn to the explicit evaluation of the Chern-Simons action for spin $N$ gauge fields describing black holes\cite{Gutperle:2011kf,Kraus:2011ds,Castro:2011fm}.

We shall not discuss the emergence of $W$ algebras (we refer to the original papers \cite{Henneaux:2010xg,Campoleoni:2010zq}) but concentrate on gauge fields leading to black hole physics. See \cite{Polyakov:1989dm} and references therein for previous work on the relation between $SL(N,\mathbb{R})$ and $W_N$ algebras. The $N=2$ leading to two copies of the Virasoro case was worked out in \cite{Coussaert:1995zp} (see also \cite{Banados:1998gg}, for a review).

The gauge fields appropriate for black hole physics have the form,
\begin{eqnarray}\label{Amu}
A_\mu = g^{-1} a_\mu g + g^{-1} \partial_\mu g\,,
\end{eqnarray}
where $g=g(r)$ is a purely radial gauge transformation, and $a_\mu$ are constant fields, in particular $a_r=0$. In components, (\ref{Amu}) reduce to,
\begin{equation}\label{ataphi}
 A_t = g^{-1}a_t g\,, \ \ \ \ \ \  A_\phi = g^{-1}a_\phi g\,, \ \ \ \ \ \ A_r = g^{-1}\partial_r g.
\end{equation}
Our first point is to show that the radial coordinate plays no role at all. In particular the action and its variation are finite without needing regularization.

We start by replacing (\ref{ataphi}) in (\ref{Ios}) and conclude that, due to the trace, the on-shell action is proportional to $\mbox{Tr}(a_t a_\phi)$. The group element $g(r)$ has dropped out. Next, we prove that the variation (\ref{dI0c}) does not depend on $g(r)$ either, even if $g(r)$ has non-zero variations at the boundary (i.e. even if radial re-parameterizations involving the varied parameters are considered). In fact, replacing (\ref{Amu}) into (\ref{dI0c}) one finds
\begin{eqnarray}\label{lt}
\delta I_0 &=& -{k \over 4\pi} \int dt d\phi\, \mbox{Tr} (  a_t\delta a_\phi - a_\phi \delta a_t + 2[a_t,a_\phi] \delta g g^{-1} )\,.
\end{eqnarray}
The first two terms (which are non-zero) do not depend on $g(r)$. The last term vanishes as a consequence of the equations $[a_t,a_\phi]=-\partial_t a_\phi+ \partial_\phi a_t=0$ for constant fields. Even if the fields were non-constant, the last term in (\ref{lt}) is zero anyway under the integral sign because $g(r)$ is only a function of $r$, and $a_t,a_\phi$ must be periodic functions on the torus.

We conclude that, as far as the on-shell action and its variations is concerned, the radial dependence plays no role
and we can simply work with the connections $a_t$ and $a_\phi$.

We shall concentrate here on non-rotating solutions for which the fields $a_\mu$ and $\bar a_\mu$ are related. We consider the subset of fields satisfying
\begin{equation}\label{j=0}
\bar a_t = a_t^T, \ \ \ \ \  \bar a_\phi = - a_\phi^T.
\end{equation}
For BTZ black holes ($N=2$) this leads to non-rotating solutions. For this class of fields the total on-shell Chern-Simons action $I_0[A]-I_0[\bar A]$ becomes,
\begin{eqnarray}\label{Ios2}
I_{0}^{\rm on-shell} &=& -{k \over 4\pi} \int dtd\phi \mbox{Tr}(a_t a_\phi)
+ {k \over 4\pi} \int dtd\phi \mbox{Tr}(\bar a_t\bar a_\phi) \nonumber\\
&=& -{k \over 2\pi} \int dtd\phi \mbox{Tr}(a_t a_\phi)
\end{eqnarray}
where we have used (\ref{Ios}).

Let us now evaluate this action on interesting solutions. On the solid torus $\phi$ and $t$ parameterize,
respectively, the non-contractible and contractible loops on the solid torus. Interesting solutions are then fields $a_t,a_\phi$ satisfying
\begin{equation}\label{holN}
Pe^{\oint a_\phi d\phi} \neq 1, \ \ \ \ \ \ \   Pe^{\oint a_t dt} =1.
\end{equation}
The first condition states that the solutions are non-trivial (cannot be brought to zero via a regular gauge transformation) while the second condition states that they are regular on the whole solid torus.

Let $a_\phi$ be a connection with non-trivial holonomy, that is, satisfying the first equation in (\ref{holN}). All invariant information in $a_\phi$ is contained in the $N-1$ Casimirs,
\begin{equation}\label{QN}
Q_2 = {1 \over 2}\mbox{Tr}(a_\phi^2), \ \ \ \ \  Q_3 = {1 \over 3}\mbox{Tr}(a_\phi^3), \ \ \ \ldots \ \ \  , Q_N = {1 \over N}\mbox{Tr}(a_\phi^{N}).
\end{equation}
We assume that all $N-1$ Casimirs are independent and can be varied independently in the action principle.

Given $a_\phi$, the time component of the gauge field, $a_t$, becomes constrained by the equations of motion
\begin{equation}\label{eqncs}
[a_\phi,a_t]=0.
\end{equation}
The general solution of (\ref{eqncs}) is an arbitrary traceless function of $a_\phi$:  $a_t=f(a_\phi)$. Due to the Cayley-Hamilton theorem, the most general traceless function is
\begin{equation}\label{atN}
a_{t} = \sigma_2 a_\phi + \sigma_3 \left( a_\phi^2 - {I \over N}\mbox{Tr}(a_\phi^2) \right) + \cdots + \sigma_{N} \left( a_\phi^{N-1} - {I \over N}\mbox{Tr}(a_\phi^{N-1}) \right)
\end{equation}
where $\sigma_2,...\sigma_N$ are $N-1$ arbitrary parameters, and $I$ represents the $N\times N$ identity matrix. The solutions $a_t,a_\phi$ are then characterized by $2(N-1)$ parameters $Q_2,Q_3,...,Q_{N}$ and $\sigma_2,\sigma_3,...,\sigma_{N}$. We know show that these parameters form conjugate pairs, and that the action can be written in a closed form in terms of them.

First note that inserting (\ref{atN}) into (\ref{Ios2}) and using the definitions (\ref{QN}), the on-shell value of $I_0$ is simply,
\begin{equation}\label{IosN}
I_0^{\rm on-shell} = -{k}\sum_{n=2}^{N} n\,\sigma_n\, Q_n.
\end{equation}
However, $I_0$ does not have the desired canonical structure at infinity and needs to be supplemented by a boundary term.

To see this we consider the action variation (\ref{dI0c}) and insert the set of solutions $a_\phi,a_t$ considered here. The bulk piece vanishes (the field is a solution) while the boundary term can be expanded as
\begin{eqnarray}
 \delta I_0 &=&   {k}\sum_{n=2}^{N} \mbox{Tr} \left( a_\phi \delta (\sigma_n a_\phi^{n-1}) - \sigma_n a_\phi^{n-1} \delta a_\phi \right) \nonumber\\
   &=&  {k} \sum_{n=2}^{N}\mbox{Tr} \left( a_\phi^{n} \delta \sigma_n + {n-2 \over n}\sigma_n \delta( a_\phi^{n}) \right)  \nonumber\\
   &=&  {2\,k} \sum_{n=2}^{N} {1\over n}\mbox{Tr}( a_\phi^{n}) \delta \sigma_n + \delta  \left( {k} \sum_{n=2}^{N} {n-2 \over n} \sigma_n\mbox{Tr}( a_\phi^n) \right) \nonumber\\
   &=& 2\, k \sum^{N}_{n=2 }  Q_n \delta \sigma_n + \delta \left( \sum^{N}_{\ n=2} {k(n-2)} \sigma_n Q_n \right)\,. \label{dq}
  \end{eqnarray}
In the last line we have used (\ref{QN}). In order to have an action whose solutions define extrema we need to make sure that the boundary term vanishes by keeping at most half of the parameters fixed.

The first piece in (\ref{dq}) vanishes if all $\sigma$'s are held fixed. But the second piece would require an extra condition on the $Q$'s. However, note that the second term in (\ref{dq}) is a total variation and hence it can be passed to the other side. In other words one considers the new action
\begin{eqnarray}\label{I1N}
I_1 &=& I_0 -  \sum^{N}_{\ n=2} {k(n-2)} \sigma_n Q_n
\end{eqnarray}
where $I_0$ is the original Chern-Simons action. From (\ref{dq}) we find that the action $I_1$ satisfies
\begin{equation}\label{di1}
\delta I_1 = \mbox{(equations of motion)} + 2\, k \sum^{N}_{\ n=2}  Q_n \delta\sigma_n
\end{equation}
and hence it has an extrema when all $\sigma$'s are held fixed. Eq. (\ref{di1}) also shows that $\{2\,k\,Q_n,\sigma_n\}$ form conjugate pairs. An action which is appropriate for fixed $Q$'s can easily be built via a Legendre transform.

The on-shell value of $I_1$ is now found from (\ref{I1N}) and (\ref{IosN}),
\begin{equation}\label{os1}
I_1^{\rm on-shell} = -k \sum^{N}_{ n=2} (2n-2)\,\sigma_n\,Q_n
\end{equation}
This is a general formula for the on-shell action, and the free energy.

The on-shell values of the action make sense only on the regular fields which satisfy the holonomy conditions (second equation in (\ref{holN})). These are $N-1$ conditions that allow to express the $N-1$ Casimirs $Q_n$, as functions of the chemical potentials $\sigma_n$. Using these relations the on-shell action $I_1^{\rm on-shell}$ can be  written as a functional of the $\sigma$'s which satisfies
\begin{equation}\label{}
{\delta I_{1}^{\rm on-shell} \over \delta \sigma_n} = 2\,k\,Q_n,
\end{equation}
consistent with (\ref{di1}). The action $I_1[\sigma_n]$ could be called `canonical'  because the charges $Q_n$ are allowed to vary, while the chemical potentials are fixed.

If one is interested in a `micro-canonical' action appropriate for fixed charges $Q_n$, instead of fixed chemical potentials, one performs a Legendre transformation to arrive at
\begin{eqnarray}
I_2&=&I_0-k\,\sum_{n=2}^N n\,\sigma_n\,Q_n\nonumber\\
&\overset{\rm on-shell}{=}&-2\,k\sum_{n=2}^N n\,\sigma_n\,Q_n
\end{eqnarray}
with
\begin{equation}
\delta I_2=-2\, k\,\sum_{n=2}^N\sigma_n\,\delta Q_n\,.
\end{equation}

To determine the energy of this set of solutions we need to identify a parameter $\beta$ which characterizes the time period. This is easily done. Consider (\ref{atN}) and redefine the chemical potentials $\sigma_n$ as
\begin{equation}\label{beta}
\sigma_n = \beta \alpha_n \ \ \ \ \mbox{with} \ \ \ \   \sum_n \alpha_n^2 =1\,.
\end{equation}
In this way, we have split the parameters $\sigma_n$ into a radius $\beta$, and a directional vector of unit norm $\alpha_n$. The motivation to do so is that $\beta$ enters as the time period. Indeed $a_t$ given in (\ref{atN}) now reads,
\begin{equation}\label{}
a_t = \beta \sum_n (...)
\end{equation}
showing that $\beta$ could be absorbed via $t\rightarrow t\beta$. (We do not make this transformation: the period of the time coordinate is fixed to $0<t<1$, and keep $\beta$ explicitly in the field.)

The variation of the action $I_1$ now reads
\begin{equation}\label{}
\delta I_1 = \mbox{(e.o.m)} + 2\,k \sum_n Q_n \alpha_n\delta \beta + 2\,k\,\beta\sum _n Q_n \delta\alpha_n,
\end{equation}
where we stress that the variations in the second term are subject to the constraint in (\ref{beta}).  The coefficient of $\delta \beta$ can now be identified with (minus) the energy of the solutions:
\begin{equation}\label{}
E =-2\, k\, \sum_n Q_n \alpha_n \,.
\end{equation}

\section{$N=3$ as an explicit example}

In this and the next section we shall work with the special case $N=3$.
An explicit representation for $a_\phi$ is
\begin{equation}\label{atN=3}
a_\phi=
\begin{pmatrix}
0 & {1\over 2}Q_2 & Q_3 \\ 1 & 0 & {1\over 2}Q_2 \\ 0 & 1 & 0
\end{pmatrix}
\end{equation}
which satisfies \eqref{QN}. $a_t$ is as in \eqref{atN} with $N=3$
and $\bar a_\phi$ and $\bar a_t$ as in \eqref{j=0}.
We take $Q_i$ and $\sigma_i$ constant.

The action $I_0$ is
\begin{equation}
I_0=-k\,(2\, \sigma_2\,Q_2+3\,\sigma_3\,Q_3)\,.
\end{equation}
The holonomy condition on $a_t$ can be interpreted as a condition on its
eigenvalues \cite{Gutperle:2011kf}.
This leads to $\det(a_t)=0$ and ${\rm Tr}(a_t^2)=8 \pi^2$ which read,
\begin{eqnarray}\label{hol}
0&=&-2\,\sigma_3^3\,Q_2^3+18\, \sigma_2^2\,\sigma_3\,Q_2^2+27\,\sigma_2\,\sigma_3^2\,Q_2\,Q_3
+27\,\sigma_3^3\,Q_3^2+27\,\sigma_2^3\,Q_3\nonumber\\
8\pi^2 & = & 2\, \sigma_2^2\, Q_2+6\,\sigma_2\, \sigma_3\, Q_3+\frac{2}{3} \sigma_3^2\, Q_2^2
\end{eqnarray}
These two conditions can be solved to express $Q_i(\sigma_j)$. This means that for given $\sigma_i$, one can find values $Q_i$ such that the gauge field is regular in the entire solid torus.

The holonomy conditions pass an important consistency check \cite{Gutperle:2011kf}: they lead to the identification of
$(Q_i,\sigma_i)$ as canonical pairs.
Assuming that $Q_i=Q_i(\sigma_j)$ we take the derivatives of the two equations in (\ref{hol}) with respect to $\sigma_j$.
This gives four linear equations for the derivatives which are solved by,
\begin{eqnarray}
 {\partial Q_2 \over\partial \sigma_2} &=& {6\over N}\,\Big(\sigma_2\,Q_2-3\,\sigma_3\,Q_3\Big)\,,\nonumber\\
 {\partial Q_2 \over\partial \sigma_3} &=& {1 \over N}\,\Big(9\,\sigma_2\,Q_3-4\,\sigma_3\,Q_2^2\Big)\,, \nonumber\\
 {\partial Q_3 \over\partial \sigma_2} &=&{1\over N}\,\Big(9\,\sigma_2\,Q_3-4\,\sigma_3\,Q_2^2\Big)\,, \nonumber\\
 {\partial Q_3 \over\partial \sigma_3}&=& {2\,Q_2\over N}\,\Big(\sigma_2\,Q_2-3\,\sigma_3\,Q_3\Big)\,,\label{deri}
\end{eqnarray}
where
\begin{equation}
N=4\,\sigma_3^2\,Q_2-3\,\sigma_2^2\,.
\end{equation}
We observe the following integrability condition\cite{Gutperle:2011kf}
\begin{equation}\label{ic}
{\partial Q_2\over\partial \sigma_3} = {\partial Q_3 \over\partial \sigma_2}.
\end{equation}
This implies that there exists a functional of $\sigma_i$ whose gradient gives $Q_i$.
From the discussion in the previous section it follows immediately that
this functional is precisely the on-shell Chern-Simons action
\begin{equation}\label{I1N=3}
I_1 =-k(2\,\sigma_2 Q_2+4 \sigma_3\,Q_3)\,.
\end{equation}
Using \eqref{I1N=3} and \eqref{deri} we also verify
\begin{equation}
{\partial I_1\over\partial \sigma_2}=2\,k\,Q_2\qquad\hbox{and}\qquad
{\partial I_1\over\partial \sigma_3}=2\,k\,Q_3
\end{equation}
directly.

Alternatively, we can use $Q_i$ as independent variables and solve
(\ref{hol}) for $\sigma_i$ in terms of $Q_i$.
This leads to
\begin{eqnarray}
{\partial\sigma_2\over\partial Q_2}&=&
-{2\,Q_2\over N}\,\Big(\sigma_2\,Q_2-3\sigma_3\,Q_3\Big)\,,\nonumber\\
{\partial\sigma_2\over\partial Q_3}&=&
\phantom{-}{1\over N}\,\Big(9\,\sigma_2\,Q_3-4\,\sigma_3\,Q_2^2\Big)\,,\nonumber\\
{\partial\sigma_3\over\partial Q_2}&=&
\phantom{-}{1\over N}\,\Big(9\,\sigma_2\,Q_3-4\,\sigma_3\,Q_2^2\Big)\,,\nonumber\\
{\partial\sigma_3\over\partial Q_3}&=&
-{6\over N}\,\Big(\sigma_2\,Q_2-3\,\sigma_3\,Q_3\Big)\,,
\end{eqnarray}
with
\begin{equation}
N=4\,Q_2^3-27\,Q_3^2\,.
\end{equation}
We conclude that the integrability condition
\begin{equation}\label{intS}
{\partial\sigma_2\over\partial Q_3}={\partial \sigma_3\over\partial Q_2}
\end{equation}
is satisfied and one also verifies
\begin{equation}
{\partial I_2\over \partial Q_2}=-2\, k\,\sigma_2\qquad\hbox{and}\qquad
{\partial I_2\over\partial Q_3}=-2\,k\,\sigma_3\,.
\end{equation}

\section{Connection with the metric formulation}
\label{R}

We have shown that the Chern-Simons action for $SL(N,\mathbb{R})$ fields can be evaluated yielding finite and sensible results.
A salient property of the on-shell action is the fact that it does not depend on the radial coordinate at all. Hence the free energy and most thermodynamical properties  can be studied at the purely topological level without ever writing an explicit metric. Of course it is interesting to write a metric, if one is interested in studying, for example, the motion of test particles.

The metric carries its own regularity conditions, namely the correct Hawking temperature, and we would like to make sure that the holonomy conditions imposed before are in one-to-one correspondence with the Hawking temperature. We shall analyze this problem in detail in a forthcoming publication. Here we highlight a method to build black holes satisfying the correct regularity conditions for $N=3$.
See \cite{Gutperle:2011kf,Castro:2011fm}, and in particular, \cite{Ammon:2011nk} for related discussions.

We introduce the radial coordinate via a gauge transformation as in (\ref{Amu}). There are two separate issues one has to take into account when introducing the radial dependence. First, as mentioned before, the Hawking temperature must have the
right values. But, even before worrying about the temperature, the differential $dt$, which becomes singular at the horizon, must appear in the physical fields multiplied by a function of $r$ that vanishes at least as $(r-r_+)$ at the horizon. (We choose the radial coordinate $r$ such that  $g_{rr}$ is constant.)  This step is automatic in the $SL(2,\mathbb{R})$ case, but as first observed in \cite{Gutperle:2011kf}, it provides non-trivial regularity conditions in the $SL(N,\mathbb{R})$ case, for $N\geq 3$.

The lesson that followed from the analysis of \cite{Gutperle:2011kf,Ammon:2011nk} is that achieving the right order of the vanishing of the functions which multiply $(dt)^n$, e.g. the double zero in $g_{tt}$,
must not be understood as a condition on the parameters $Q_i$ and $\sigma_i$. Instead, this is a condition on the gauge transformation $g(r)$ when introducing the radial coordinate. The parameters $\sigma_i$ are still determined by holonomy conditions, or by Hawking conditions, which should be the same.

The components of the $sl(N)$ dreibein are
\begin{eqnarray}
  e_t &=& g_{1}^{-1}\, a_t\, g_1 - g_{2}^{-1}\,\bar a_t\, g_2\nonumber \\
  e_\phi &=& g_{1}^{-1}\, a_\phi\, g_1- g_{2}^{-1}\, \bar a_\phi\, g_2 \\
  e_r &=& g_{1}^{-1}\partial_r g_1 - g_2^{-1} \partial_r g_2\nonumber \label{e}
\end{eqnarray}
where $\bar a$ is the connection of the second $SL(N)$ factor and we allow for
two different group elements.

We now specify to the case $N=3$ which contains the metric and a
spin three field. As shown in \cite{Campoleoni:2010zq},
in terms of the dreibein, the metric $g_{\mu\nu}$ and spin three field $\psi_{\mu\nu\rho}$
are
\begin{eqnarray}\label{metrics}
g_{\mu\nu}dx^\mu dx^\nu &=& \mbox{Tr} \big(( e_t\,dt + e_\phi\,d\phi + e_r\,dr )^2 \big),\nonumber  \\
\psi_{\mu\nu\rho} dx^\mu dx^\nu dx^\rho &=& \mbox{Tr} \big(( e_t\,dt + e_\phi\,d\phi + e_r\,dr )^3 \big).
\end{eqnarray}

In order for the metric and spin three field  to be regular at the horizon, we shall define the (non-extremal, non-rotating) horizon as a point where the time component of the spin 3 dreibein vanishes linearly
\begin{equation}\label{eh}
e_t(r_+) =0\,,\qquad \partial_r e_t(r_+)\neq 0\,.
\end{equation}
In this way, at the horizon $g_{tt}$ will vanish quadratically, while the spin 3 component $\psi_{ttt}$ will vanish cubically. More generally, from (\ref{metrics}), we observe that (\ref{eh}) implies that any component of any of the two fields carrying $dt$, will vanish at the horizon with the correct power. We stress that (\ref{eh}) only applies to non-rotating black holes. The general case will be described elsewhere.

The definition of the horizon (\ref{eh}) can be translated into a condition on the group elements. From (\ref{e}) we find
\begin{eqnarray}\label{}
e_t = g_1^{-1} (a_t\, U(r) - U(r)\, \bar a_t)\, g_2.
\end{eqnarray}
where
\begin{equation}\label{U}
U(r) = g_1^{\phantom{|}} g_2^{-1}.
\end{equation}
Incidentally, we mention that all components of the metric only depend on $U(r)$, and not on the individual group elements $g_1$ and $g_2$.

The horizon can now be defined as a point $r_0$ where the group element (\ref{U}) satisfies,
\begin{equation}\label{U0cond}
a_t\, U_0 = U_0\, \bar a_t, \ \ \ \ \ U_0 \equiv U(r_+)\,.
\end{equation}
The connections $a_t,\bar a_t$ depend on the charges $Q_i,\,\sigma_i$. This condition then expresses the horizon radius in terms of the charges.  For our choice, \eqref{atN=3} and \eqref{j=0}, there is a three-parameter
solution for $U_0$ which satisfies \eqref{U0cond}, the simplest being
\begin{equation}\label{U0}
U_0 =  \left(\begin{array}{ccc}
        0 & 0 & -1 \\
        0 & -1 & 0 \\
        -1 & 0 & 0
      \end{array} \right)\,.
\end{equation}
The group element $U(r)$ is largely arbitrary except that at some point $r_+$ it has the value (\ref{U0}). In particular, one could take $U(r)$ to have the form ,
\begin{equation}\label{}
U(r) =
\begin{cases}
e^{2 r L} & \mbox{for} ~~~ r\to\infty\,,\\
U_0 & \mbox{for}~~~ r\to r_+\,.
\end{cases}
\end{equation}
where $L$ is a constant $sl(3,\mathbb{R})$ element.
In this way, the asymptotics take the usual form and the horizon is regular.

Let us consider a concrete particular case.
For non-rotating black holes the only non-vanishing components of $g$ and $\psi$ are
\begin{eqnarray}\label{gpsi}
  g_{\mu\nu}dx^\mu dx^\nu &=& - g_{tt}\, dt^2 + g_{rr}\, dr^2 + g_{\phi\phi}\,d\phi^2\,,\nonumber \\
   \psi_{\mu\nu\rho}dx^\mu dx^\nu dx^\rho &=& d\phi\,(- \psi_{\phi tt}\, dt^2 + \psi_{\phi rr}\, dr^2 + \psi_{\phi\phi\phi}\,d\phi^2 )\,.
\end{eqnarray}
Since $d\phi$ is a regular coordinate on the whole solid torus, both fields represent black holes in the usual sense.
Having defined the radial dependence via $g_1$ and $g_2$,
we automatically have the right order of zeros multiplying $dt$.
We stress, once more, that at this point these conditions impose no restriction at all on the parameters $Q_i,\,\sigma_i$. Restrictions arise from the requirement that there is no
conical singularity at the horizon, both for the metric $g$ and the spin-3 field $\psi$. These conditions are, for the case under
consideration,
\begin{equation}\label{regcond}
{-g_{tt}\over r^2 g_{rr}}\Big|_{r=r_0}=4 \pi^2\quad\hbox{and}\quad
{-\psi_{\phi tt}\over r^2\psi_{\phi rr}}\Big|_{r=r_0}=4\pi^2
\end{equation}

Imposing the particular form \eqref{gpsi} puts additional restrictions on $U$.
To find a regular solution we make the Ansatz
\begin{equation}
g_1=g_{10}\, e^{r J}\,,\qquad g_2=g_{20}\,e^{-r J}\qquad\hbox{with}\quad J\in sl(3)
\end{equation}
and
\begin{equation}
g_{10}=I\,,\qquad g_{20}=U_0\,.
\end{equation}
such that
\begin{equation}
U=U_0\, e^{2 r J}\,.
\end{equation}
Note that $U(r)\rightarrow U_0$, as $r\rightarrow 0$. The horizon is thus located at $r=0$.

We also require that the limit $Q_2\to0,\,\sigma_3\to 0$ is smooth and leads to the BTZ metric
and $\psi=0$.
With some effort one shows that the following $J$ satisfies all the above requirements,
\begin{equation}
J=
\begin{pmatrix}
1 & 0 & 0 \\ 0 & 0 & 0 \\ 0 & 0 & -1
\end{pmatrix}
-
\begin{pmatrix}
0 & s & 0 \\ 0 & 0 & -s \\  0 & 0 & 0
\end{pmatrix}
\end{equation}
with
\begin{equation}
s={Q_3\over Q_2}\sum_{n=0}^\infty{1\over 2n+1}\binom{3 n}{n}\left({Q_3^2\over Q_2^3}\right)^n
={2\over\sqrt{3}}\sqrt{Q_2}\,\sin\left({1\over3}\sin^{-1}\left({3\over2}\sqrt{3 Q_3^2\over Q_2^3}\right)\right)
\end{equation}
One can also check that the metric and spin-3 components are even functions of $r$ \cite{Ammon:2011nk}
and that in the BTZ limit we can identify $Q_2=M$ and $\sigma_2=\beta$.\footnote{We have
not checked that this solution is the same as the one constructed in \cite{Ammon:2011nk}. Due to the
use of different radial variables and different parametrization of the
connection,
this check is not completely trivial.}
The details will be presented elsewhere.

If one inserts the resulting metric and spin-3 components into the regularity conditions
(\ref{regcond}), one can use them to express the $Q_i$ in terms of the $\sigma_i$. The first few terms
in a power series expansion in $\sigma_3$ are
\begin{eqnarray}
Q_2&=&{4\,\pi^2\over\sigma_2^2}+{80\,\pi^4\sigma_3^2\over 3\,\sigma_2^6}+{640\,\pi^6\sigma_3^4\over 3\,\sigma_2^{10}}
+{56576\,\pi^8\sigma_3^6\over 27\, \sigma_2^{14}}+{1845248\,\pi^{10}\sigma_3^8\over 81\, \sigma_2^{18}}+\dots\\
Q_3&=&-{32\,\pi^4\sigma_3\over 3\,\sigma_2^5}-{2560\,\pi^6\sigma_3^3\over 27\,\sigma_2^9}-{8704\,\pi^8\sigma_3^5\over 9\,\sigma_2^{13}}
-{868352\,\pi^{10}\sigma_3^7\over 81\,\sigma_2^{17}}+\dots
\end{eqnarray}
One checks that this also solves the holonomy constraints (\ref{hol}).

We close by mentioning possible generalizations of our analysis.
The generalization to rotating solutions is straightforward. In this
case the parameters in $a$ and $\bar a$ are unrelated. One can also discuss extremal spin-3 black holes. Another issue which we shall discuss elsewhere is the possibility of phase transitions between the different branches of the regularity conditions:  the conditions which express $Q_i$ as functions of $\sigma_i$ are non-linear and have several branches. It may happen that the free energy is a minimum for different branches in different regions of the space of parameters. This would be a signal for phase transitions.

\bigskip

\noindent
{\bf Acknowledgements}

We are grateful to A. Castro and P. Kraus for a detailed reading of and useful comments on an earlier version of
the manuscript. The work of MB was partially supported by Fondecyt (Chile) Grants \#1100282 and \# 1090753. The work of RC is supported by a PhD Conicyt (Chile) grant \# 21090235. ST thanks the Departamento de F\'\i sica at Universidad Cat\'olica in Santiago for hospitality while this project was started
and the Physics Department of the University of Heidelberg while it was finished.
He also acknowledges generous support from the Klaus Tschira Foundation.

\appendix

\setcounter{equation}{0}

\appendix

\setcounter{equation}{0}

\section{Appendix}

In most of the literature a different parametrization of the gauge connections is used.
The chiral structure of the asymptotic symmetry algebra -- two copies of the
$W_N$ alegebra -- is most easily discussed in this parametrization. The relation between the
parametrization of this appendix with the one used in the main body of the text is
easily worked out explicitely.

The simplest (non-zero) solution to the Chern-Simons equations of motion is the chiral field with $a_{\bar z}=0$ and $a_z = a_z(z)$.  This solution, supplemented with $a_r=0$, satisfies $F_{\mu\nu}=0$ and leads to an asymptotic affine symmetry (for a review see, for example, \cite{Banados:1998gg}). However, in order to describe AdS asymptotics extra conditions on the currents need to be incorporated. This was first pointed out in the \cite{Coussaert:1995zp} in the context of the
$SL(2,\mathbb{R})\times SL(2,\mathbb{R})$ theory, following earlier work on the Hamiltonian reduction of WZW theories. For the higher spin theory, the extra boundary conditions were discussed in \cite{Henneaux:2010xg,Campoleoni:2010zq}. See \cite{Polyakov:1989dm} and references therein for previous work on the relation between $SL(N,\mathbb{R})$ and $W_N$ algebras.

Following these works we consider $N=3$ with currents
\begin{equation}\label{az}
  a_{z} = \left( \begin{array}{ccc}
                 0 & {1\over 2}T & W \\
                 1 & 0 & {1\over 2}T \\
                 0 & 1 & 0
               \end{array} \right), \ \ \ \  a_{\bar z}=0,
\end{equation}
where $T$ and $W$ are functions of $z$ only and
$T = {1 \over 2} \mbox{Tr}(a_z^2)\,, W = {1 \over 3} \mbox{Tr}(a_z^3)$.
In the black hole context $T$ and $W$ are constants.
Gauge fields of the form (\ref{az}) are a natural generalization of the Coussaert-van Driel-Henneaux \cite{Coussaert:1995zp} construction of the Virasoro symmetry in the $sl(2)$ theory. The transformation of $T$ and $W$ under the $W_3$ generated by them can be obtained straightforwardly from considering the $sl(3)$ transformations with parameters $\lambda$,
\begin{equation}\label{}
\delta a_z = \partial_z \lambda + [a_z,\lambda], \ \ \ \ \  \lambda \in sl(3,\mathbb{R})
\end{equation}
such that they leave the form (\ref{az}) invariant \cite{Polyakov:1989dm,Campoleoni:2010zq}.   This is the Brown-Henneaux (or Regge-Teitelboim) representation of the conformal symmetry. In the AdS/CFT formulation, the conformal properties are derived  by turning on sources ($a_{\bar z}\neq 0$) and interpreting the action as a generating functional. The conformal Ward identities are derived by usual methods. See \cite{Gutperle:2011kf} for the explicit calculation in the spin 3 case. An earlier calculation in the spin 2 case can be found in \cite{Banados:2004nr}.

The general situation would be to introduce, for the second
$SL(3)$ factor,  the connection $\bar a$ with two new charges.
The simplest solution will be if they are related to the charges in $a$
via the choice $\bar a_{\bar z} = a_z^T$.

The fields (\ref{az}) and their barred counterparts are perfectly well-behaved as asymptotic fields, but they cannot be extended on the whole solid torus. In order to find well-behaved fields on the solid torus we need to consider a more general set of fields. Of course, we would like to keep track of the conformal properties.

This problem is not exclusive of the higher spin solutions, but it is also present in the BTZ metric. Let us briefly recall this issue in the case of the rotating BTZ metric. The rotating BTZ metric contains a term $N^\phi(r) dt$ which is singular at the horizon $r=r_+$ unless $N^\phi(r_+)=0$. On the other hand, the Brown-Henneaux conditions require $N^\phi(r) \simeq 0$ for large $r$. Both choices are related by the coordinate transformation\footnote{See \cite{Gutperle:2011kf} for a recent related discussion concerning charged black holes.},
\begin{equation}\label{av}
\phi \rightarrow \phi - \Omega t,
\end{equation}
but they cannot be satisfied simultaneously. The regular metric with $N^\phi(r_+)=0$ has the large $r$ behavior,
\begin{equation}\label{}
ds^2 \simeq r^2(-dt^2 + d\phi^2) + {dr^2 \over r^2} + r^2\Omega dt(2 d\phi + \Omega dt),
\end{equation}
differing from the standard Brown-Henneaux fall-off conditions. Now, this does not mean that the conformal symmetry is broken. One only has to remember to undo the shift (\ref{av}) when computing the charges.

For black holes with spin 3 charge we encounter a similar problem. The connections (\ref{az}) do not generate a smooth black hole all the way in the interior of the torus. What is missing are the chemical potentials associated to the charges $T,W$, just like the angular velocity $\Omega$ is the chemical potential associated to $J$.

In full analogy with (\ref{av}), the extension we need can be found via a simple gauge transformation. Let $X$ be a constant Lie algebra-valued matrix which commutes with $a_z$ (given in (\ref{az})) and consider a gauge transformation with group element
\begin{equation}\label{UX}
U(\bar z) = e^{\bar z X}.
\end{equation}
Since $[a_z,X]=0$ the new fields are simply,
\begin{equation}\label{primed}
a'_z = a_z, \ \ \ \  a'_{\bar z} = X, \ \ \ \ a'_r =0\,.
\end{equation}
The role of this gauge transformation is to turn on $a_{\bar z}$. We drop the primes and consider the set of connections (\ref{az}) and a non-zero $a_{\bar z}$ satisfying
\begin{equation}\label{}
[a_z,a_{\bar z}]=0.
\end{equation}
with the general solution
\begin{equation}\label{azb}
a_{\bar z} = \mu\, a_z + \nu \Big( a_z^2 - {1 \over 3} \mbox{Tr}(a_z^2) \Big)\,;
\end{equation}
$\mu,\nu$ are arbitrary parameters. The new space of solutions is then given by the connection $a_z$ in (\ref{az}) and
$a_{\bar z}$ in (\ref{azb}), characterized by the four parameters $T,W,\mu,\nu$. The new parameters $\mu,\nu$ allow to build a regular connection that can be extended to the interior of the solid torus.

The fields $a_z$ (\ref{az})  and $a_{\bar z}$ (\ref{azb}) exhaust the set of constant flat connections with non-zero quadratic and cubic Casimirs and contains all examples considered in the literature so far (e.g. \cite{Gutperle:2011kf,Ammon:2011nk,Castro:2011fm}), up to redefinitions and similarity transformations.
It turns out, however, that the parametrization in terms of $a_\phi$ and $a_t$ used in the main text is more convenient for the
discussion of the canonical structure of black hole solutions.

The gauge transformation (\ref{UX}) is analogous to the coordinate transformation (\ref{av}).  In the spin 2 case, one could have chosen to introduce $\Omega$ via a transformation of the form (\ref{UX}) instead of the coordinate redefinition (\ref{av}). The final result is the same. In the $sl(3)$ case, it is easier to deal with the transformation (\ref{UX}) instead of a coordinate redefinition followed by a spin 3 transformation.

It is important to note that the group element $U$ is not trivial, in fact, it is not periodic on the torus (with a cycle defined by $0\leq t < 1$),
\begin{equation}\label{hU}
U(t=0) \neq U(t=1).
\end{equation}
As a consequence the holonomies of the gauge transformed field do not coincide with those of the original field with
$a_{\bar z}=0$. This is expected and correct. We have started with a singular connection (\ref{az})  and act on it with a singular gauge transformation. The outcome is a regular field. This is very much similar to the Kruskal coordinates for black holes. The Schwarzschild coordinates are singular at the horizon while the Kruskal coordinates are regular. This necessarily means that the relation between both set of coordinates is singular. Once the regular field is reached, one should, in principle, forget about the irregular one. The reason one insists in keeping the irregular one is because the asymptotic CFT symmetry, defined at infinity, has a simple form. Again, this is similar to Schwarzschild black holes. Asymptotically, Schwarzschild coordinates are more useful  than Kruskal ones.


\end{document}